\input phyzzx.TEX
\tolerance=1000
\voffset=-0.0cm
\hoffset=0.7cm
\sequentialequations
\def\rl{\rightline}

\def\t1{{\tilde 1}}

\def\t{\theta}

\REF{\BEK}{J. Bekenstein, Lett. Nuov. Cimento {\bf 4} (1972) 737; Phys Rev. {\bf D7} (1973) 2333; Phys. Rev. {\bf D9} (1974) 3292;
S. Hawking, Nature {\bf 248} (1974) 30; Comm. Math. Phys. {\bf 43} (1975) 199.}
\REF{\LAST}{E. Halyo, [arXiv:1403.2333[hep-th]].}
\REF{\LEN}{L. Susskind, [arXiv:hep-th/9309145].}
\REF{\WAL}{R. M. Wald, Phys. Rev. {\bf D48} (1993) 3427, [arXiv:gr-gc/9307038]; V. Iyer and R. M. Wald, Phys. Rev. {\bf D50} (1994) 
846, [arXiv:gr-qc/9403028]; Phys. Rev. {\bf D52} (1995) 4430, [arXiv:gr-qc/9503052].}
\REF{\HOL}{G. 't Hooft, [arXiv:gr-qc/9310026]; L. Susskind, J. Math. Phys. {\bf 36} (1995) 6377, [arXiv:hep-th/9409089]; R. Bousso, Rev. Mod. Phys. {\bf 74} (2002) 825, [arXiv:hep-th/0203101].}
\REF{\REV}{T. Padmanabhan, Rep. Prog. Phys. {\bf 72} (2010) 046901, [arXiv:0911.5004[gr-qc]]; Phys. Rep. {\bf 406} (2005) 49,
[arXiv:gr-qc/0311036].}
\REF{\GIB}{G. W. Gibbons and S. W. Hawking, Phys. Rev. {\bf D15} (1977) 2752.}
\REF{\MED}{E. Halyo, JHEP {\bf 0206} (2002) 012, [arXiv:hep-th/0201174]; A. J. M. Medved, [arXiv:hep-th/0201215].}
\REF{\SBH}{E. Halyo, A. Rajaraman and L. Susskind, Phys. Lett. {\bf B392} (1997) 319, [arXiv:hep-th/9605112].}
\REF{\EDI}{E. Halyo, Int. Journ. Mod. Phys. {\bf A14} (1999) 3831, [arXiv:hep-th/9610068]; Mod. Phys. Lett. {\bf A13} (1998), [arXiv:hep-th/9611175].}
\REF{\DES}{E. Halyo, [arXiv:hep-th/0107169].}
\REF{\UNI}{E. Halyo, JHEP {\bf 0112} (2001) 005, [arXiv:hep-th/0108167].}
\REF{\DIV}{T. Padmanabhan, Int. Journ. Mod. Phys. {\bf D15} (2006) 1659 [arXiv:gr-qc/0606061].}
\REF{\SUR}{B. R. Majhi and T. Padmanabhan, Eur. Phys. Journ. {\bf C73} (2013) 2651, [arXiv:1302.1206[gr-qc]].}
\REF{\SURF}{T. Padmanabhan, AIP Conf. Proc. {\bf 1483} (2012) 212, [arXiv:1208.1375[hep-th]]; Gen. Rel. Grav. {\bf 44} (2012) 2681,
[arXiv:1205.5683[gr-qc]].}
\REF{\PGH}{T. Padmanabhan, Int. Journ. Mod. Phys. {\bf D15} (2006) 2029, [arXiv:gr-qc/0609012].}
\REF{\ADS}{J. Maldacena, Adv. Theor. Math. Phys. {\bf 2} (1998) 231, [arXiv:hep-th/9711200]; S. Gubser, I. Klebanov and A. Polyakov, Phys. Lett. {\bf B428} (1998) 105,
[arXiv:hep-th/9802109]; E. Witten, Adv. Theor. Math. Phys. {\bf 2} (1998) 253, [arXiv:hep-th/9802150].}
\REF{\BIR}{D. Birmingham, JHEP {\bf 9906} (1999) 036, [arXiv:hep-th/9906040].}
\REF{\VER}{E. Verlinde, [arXiv:hep-th/0008140]; L. Savonije and E. Verlinde, Phys. Lett. {\bf B507} (2001) 305, 
[arXiv:hep-th/0102042].} 
\REF{\LOV}{C. Lanczos, Z. Phys. {\bf 73} (1932) 147; D. Lovelock, Journ. Math. Phys. {\bf(12)} (3) (1971) 498.}
\REF{\LAN}{T. Padmanabhan and D. Kothawala, Phys. Rep. {\bf 531} (2013) 115, [arXiv:1302.2151[gr-qc]].}
\REF{\MSUR}{A. Mukhopadhyay and T. Padmanabhan, Phys. Rev. {\bf D74} (2006) 124023 [arXiv:hep-th/0608120].}
\REF{\GB}{R. G. Cai, Phys. Rev. {\bf D65} (2002) 084014, [arXiv:hep-th/0109133]; M. Cvetic, S. Nojiri and S. D. Odintsov,
Nucl. Phys. {\bf B628} (2002) 295, [arXiv:hep-th/0112045]; T. Clunan, S. F. Ross and D. J. Smith, Class. Quant. Grav. {\bf 21} (2004) 3447, [arXiv:gr-qc/0402044].}
\REF{\GRU}{P. Kraus and F. Larsen, JHEP {\bf 0601} (2006) 022, [arXiv:hep-th/0508218]; D. Grumiller, M. Irakleidou, I. Lovrekovic,
and R. McNees, Phys. Rev. Lett. {\bf112} (2014) 111102 [arXiv:1310.0819[hep-th]].}

\singlespace
\rl{SU-ITP-14/14}
\pagenumber=0
\normalspace
\medskip
\bigskip
\titlestyle{\bf{On the Holographic Nature of Rindler Energy}}
\smallskip
\author{ Edi Halyo{\footnote*{e--mail address:halyo@stanford.edu}}}
\smallskip
\centerline {Department of Physics} 
\centerline{Stanford University} 
\centerline {Stanford, CA 94305}
\smallskip
\vskip 2 cm
\titlestyle{\bf ABSTRACT}

We show that the dimensionless Rindler energy of a black hole, $E_R$, is exactly the surface Hamiltonian obtained from the Einstein--Hilbert action evaluated on the horizon. Therefore, $E_R$ is given by a surface integral over the horizon and
manifestly holographic. In the context of the AdS/CFT duality, Rindler energy corresponds, on the boundary, to a dimensionless energy given by the product of the AdS radius and the extensive part of the CFT energy. We find that, beyond General Relativity, $E_R$
is still holographic but not necessarily given by the surface Hamiltonian of the theory.

\singlespace
\vskip 0.5cm
\endpage
\normalspace

\centerline{\bf 1. Introduction}
\medskip

Recently, it was shown that the entropy of any nonextreme black hole[\BEK] in any theory of gravity is given by $S=2 \pi E_R$ where
$E_R$ is the dimensionless Rindler energy obtained in the near horizon region of the black hole[\LAST].
In fact, $E_R$ is identical to Wald's Noether charge so that the above entropy is exactly Wald entropy. The easiest way to see this
relation is to 
note that $E_R$ is the dimensionless energy near the horizon in units normalized so that $T=1/2\pi$[\LEN] whereas Wald's Noether charge is the generator of time translations near the horizon with the surface gravity normalized to be $\kappa=1$[\WAL].
 

In ref. [\LAST] some issues related to $E_R$ were left unresolved. First, the derivation of $E_R$ is equivalent to the First Law of Thermodynamics and not a new computation of entropy. Therefore, it is desirable to find a formula for $E_R$ in terms of the gravitational degrees of freedom, i.e. the metric. Second, $E_R$, just like Hawking temperature, is obtained directly from the time
evolution in the near horizon region of the black hole. As a result, its holographic nature is not manifest. It would be nice if 
$E_R$ were given by  a surface integral over the horizon clarifying its relation to holography[\HOL].
Third, the relation between time evolution and entropy is obscure. Entropy is related to the number of black hole microstates that are
compatible with the black hole metric. This implies that there must be a relation between the number of microstates of a black hole and time evolution in the near horizon region. Fourth, using the AdS/CFT correspondence, we can find the analog of $E_R$ for an AdS black hole on the boundary. We expect to find a dimensionless energy in the boundary CFT that gives the entropy of the black hole.
Finally and perhaps most importantly, we do not the nature of the fundamental degrees of freedom that are counted by $E_R$.

In this paper, we address the first four issues mentioned above in General Relativity. We show that $E_R$ is given by the surface integral over the horizon 
of an expression that depends only on the metric. This is the desired formula for $E_R$ that is manifestly holographic. It is 
well--known that the Einstein--Hilbert action can be divided in to a bulk and a surface term[\REV]. The Hamiltonian obtained from the surface action and computed near the horizon, with the normalization $\kappa=1$, gives $E_R$. 
Alternatively, $E_R$ can be expressed as the integral of the gradient of the local acceleration which can be rewritten as the surface
integral of a term that depends only on $g_{00}$[\REV]. This establishes the
relation between time evolution in the near horizon region and black hole entropy. 
The expression for $E_R$ is not covariant which is to be expected since black hole entropy is observer dependent. Thus, $E_R$ depends either on the metric seen by an observer, or equivalently, on her world--line.
We note that the surface action that gives rise to $E_R$ is different from the Gibbons--Hawking term used in Euclidean gravity[\GIB] but reduces to it in most cases of interest.
Finally, we discuss the holographic nature $E_R$ in the context of the AdS/CFT correspondence. We consider AdS black holes and show that their entropy is given by a dimensionless energy in the boundary CFT, 
which is the product of the AdS radius and the extensive part of the CFT energy. This is the holographic counterpart of $E_R$[\MED].

The paper is organized as follows. In the next section, we review the derivation and properties of the dimensionless Rindler energy $E_R$. In section 3, we obtain the holographic formula for $E_R$ in General Relativity and discuss its properties. In section 4, we consider AdS black holes and find the counterpart of $E_R$ in the boundary CFT. In section 5, we discuss the generalization of our results to theories of gravity beyond General Relativity. Section 6 contains a summary and discussion of our results.

\bigskip
\centerline{\bf 2. Dimensionless Rindler Energy as Entropy}
\medskip

For completeness, we first describe the dimensionless Rindler energy of any (nonextreme) black hole which is a review of section 2 of ref. [\LAST].
In any theory of gravity, the metric of a nonextreme black hole in D--dimensions is of the form
$$ ds^2=-f(r)~ dt^2+ f(r)^{-1} dr^2+ r^2 d \Omega^2_{D-2} \quad. \eqno(1)$$
The radius of the horizon, $r_h$, is determined by
$f(r_h)=0$. If in addition, $f^{\prime}(r_h) \not =0$, the near horizon region is described by Rindler space. Near the
horizon, $r=r_h +y$ where $y<<r_h$ and to the lowest order 
$f(r)=f(r_h)+f^{\prime}(r_h)y$ . In terms of $y$ the metric becomes
$$ds^2=-f^{\prime}(r_h)y~ dt^2+(f^{\prime}(r_h)y)^{-1} dy^2+ r_h^2 d \Omega^2_{D-2} \quad. \eqno(2)$$
Using the proper radial distance, $\rho$, obtained from $d\rho=dy/\sqrt{f^{\prime}(r_h)y}$ the metric can be written as
$$ds^2=-{f^{\prime 2}(r_h) \over 4} \rho^2 dt^2+d \rho^2+ r_h^2 d \Omega^2_{D-2} \quad. \eqno(3)$$
Defining the dimensionless Rindler time as $\tau_R=(f^{\prime}(r_h)/2)~ t$, we get the black hole metric 
$$ds^2=-\rho^2 d \tau_R^2 + d \rho^2 + r_h^2 d \Omega^2_{D-2} \quad, \eqno(4)$$
where the metric in the $\tau_R-\rho$ directions describes Rindler space.

In Ref.[\LEN] it was shown that the entropy of the black hole is given by the dimensionless Rindler energy $E_R$ conjugate to $\tau_R$. $E_R$ can be obtained from the Poisson bracket
$$1=\{E_R,\tau_R\}=\left({\partial E_R \over \partial M}{\partial \tau_R \over \partial t}-{\partial E_R \over \partial t}
{\partial \tau_R \over \partial M} \right) \eqno(5)$$
where $M$ is the mass of the black object conjugate to $t$. Assuming that Hawking radiation is negligible and 
$E_R$ is time independent (which is a good approximation for large black holes) we find
$$dE_R={2 \over f^{\prime}(r_h)}~ dM \quad.\eqno(6)$$
Since Hawking temperature is given by $T_H=f^{\prime}(r_h)/4 \pi$, eq. (6) can be written as
$$d(2 \pi E_R)={dM \over T_H} \quad. \eqno(7)$$
which is the First Law of Thermodynamics. As a result, $S=2 \pi E_R$ for any black object in any theory of gravity. 
This method works for all nonextreme black objects and space--times with Rindler--like near horizon geometries[\SBH-\UNI]. 
The entropy of extreme black holes can be obtained from the extreme limit of nonextreme black holes.

In generalized theories of gravity, black hole entropy is usually taken to be Wald entropy which is given by $S_{Wald}=2 \pi Q$
where $Q$ is Wald's Noether charge[\WAL].
The equivalence between $E_R$ and Wald's Noether charge has been shown in detail in ref. [\LAST] and will not be repeated here. 
It is sufficient to note that $E_R$ is
the dimensionless energy in Rindler space (i.e. the near horizon region) with $T=1/2 \pi$ whereas $Q$ is the generator of time translations near the horizon with the normalization $\kappa=1$. Thus, $E_R$ is exactly $Q$.

In ref. [\LAST], a number of issues related to $E_R$ were left unresolved including:

1. It is clear from eq. (7) that the Poisson bracket for $E_R$ in eq. (5) is equivalent to using the First Law of Thermodynamics, i.e. the relation $T(M)$, to obtain the entropy. It is important to find a formula for $E_R$ in terms of the metric which does not directly use thermodynamics. 

2. The holographic nature of $E_R$ is not clear. It would be nice if the formula for $E_R$ made its holographic
nature manifest, e.g. if $E_R$ were given by a surface integral over the black hole horizon just like Wald's Noether charge.

3. $E_R$ is determined only from $g_{00}$, i.e. the time evolution in the near horizon of the
black hole. For black holes, a direct relation between time evolution and and entropy is certainly worth establishing.

4. If $E_R$ is holographic, in the context of the AdS/CFT correspondence, it would be interesting to find out the dimensionless 
energy in the boundary CFT that corresponds to $E_R$.

In the following sections, we address all these issues.  We obtain a formula for $E_R$ that is a surface integral over the horizon of a quantity that depends only on the metric. This is the desired formula that is manifestly holographic.
Using a geometric identity, we show that the formula for $E_R$ can be written as a surface integral of the local acceleration which is determined only by $g_{00}$. This establishes the relation between black hole entropy and the time evolution. 
Finally, we find the analog of $E_R$ for an AdS black hole in the boundary CFT which is given by the product of the AdS radius and the extensive part of the CFT energy.

\bigskip
\centerline{\bf 3. Holography and Rindler Energy}
\medskip

In this section, we obtain a holographic formula for $E_R$ in four dimensional General Relativity. The generalization of the formula to any dimension is straightforward. We discuss the holographic nature of $E_R$ in theories of gravity beyond General Relativity in section 5.

In the following we use the ($D=4$) Euclidean Rindler metric with dimensionless Rindler time obtained from eq. (4) by $\tau_R \to i \tau_E$
$$ds^2=\rho^2 d \tau_E^2 + d \rho^2 + r_h^2 d \Omega^2_2 \quad. \eqno(8)$$
The dimensionless Euclidean Rindler time coordinate, $\tau_E$, is an angle with period $\beta=1/T_H=2 \pi$. In this metric, the horizon is at $\rho=0$.

The Einstein--Hilbert action that describes $D=4$ (Euclidean) General Relativity can be divided into a bulk and a surface part
[\REV,\DIV] 
$$A_{EH}= {1 \over {16 \pi G}} \int d^4 x \sqrt{g} R = A_{bulk}+A_{surf} \quad, \eqno(9)$$
where
$$A_{bulk}={1 \over {16 \pi G}} \int d^4 x \sqrt{g}g^{ik}(\Gamma^m_{il} \Gamma^l_{km} - \Gamma^l_{ik} \Gamma^m_{lm}) \quad, \eqno(10)$$
and
$$A_{surf}={1 \over {16 \pi G}} \int d^4 x \partial_c [\sqrt{g}(g^{ck}\Gamma^m_{km} - g^{ik} \Gamma^c_{ik})] \quad. \eqno(11)$$

$A_{bulk}$ is the Dirac--Schroedinger action that describes the dynamics of gravity in the bulk. It is quadratic in 
$\Gamma$'s, so it is given by the square of the first derivatives of the metric. The only properties of gravity that $A_{bulk}$
cannot describe are those, such as black hole entropy, that depend on the boundaries.
$A_{surf}$ is a total derivative and usually it is either set to zero or canceled by a counterterm. It contains
second derivatives of the metric and can also be written as[\SUR,\SURF]
$$A_{surf}=-{1 \over {16 \pi G}} \int_S d^3 x {1 \over {{\sqrt{g}}}} \partial_i(g g^{ij}) \quad, \eqno(12)$$
where $d^3x=dt_E~d^2x_{\perp}$ with $x_{\perp}$ denoting the directions along the boundary surface perpendicular to direction
$x^c$ in eq. (11). The integral is over the surface of the three--dimensional boundary of Euclidean space--time.
From the surface action above we can compute the surface Hamiltonian by
$$H_{surf}= -{\partial A_{surf} \over \partial t_E}= {1 \over {16 \pi G}} \int_S d^2 x_{\perp} {1 \over {{\sqrt{g}}}} 
\partial_i(g g^{ij}) \quad. \eqno(13)$$

In the presence of a black hole, the boundary of the space--time consists of two separate parts: the surface of the horizon and the surface at asymptotic infinity. The integral over the whole boundary vanishes. However, we can consider the integral
over only the horizon surface, ${\cal H}$, which we denote by $H_{surf} \vert_{\cal H}$. This is the energy in the near horizon region, i.e. the Rindler energy. 
If we use the near horizon metric in eq. (8) with dimensionless time (and $\kappa=1$), $H_{surf} \vert_{\cal H}$ becomes the dimensionless Rindler energy $E_R$. In fact, a simple calculation gives
$$H_{surf}\vert_{\cal H}= {A \over {8 \pi G}} \quad, \eqno(14)$$
where $A=\int_{\cal H}d^2x_{\perp}$ is the area of the horizon.
We see that the black hole entropy is given by $S=2 \pi H_{surf}\vert_{\cal H}$ and therefore $H_{surf}\vert_{\cal H}=E_R$ or 
$$E_R={1 \over {16 \pi G}} \int_{\cal H} d^2 x_{\perp} {1 \over {{\sqrt{g}}}} \partial_i(g g^{ij}) \quad, \eqno(15)$$
where the integration is now over the two--dimensional surface of the horizon.

Eq. (15) is the desired formula for $E_R$ in terms of the metric. It should be contrasted with 
eq. (5) which is equivalent to the First Law of Thermodynamics. Eq. (15) is also a surface integral over the horizon
making the holographic nature of Rindler energy manifest. This resolves the first two issues raised at the end of
the previous section.

We note that, just like the procedure that uses the Poisson bracket in eq. (5), the formula for $E_R$ is also off--shell.
It is clear that we need to know only the metric in order to calculate the entropy. The metric does not have to satisfy the equations
of motion. In fact, we do not even need to know the gravitational Lagrangian.

The entropy of the black hole is also given by an alternative expression in terms of the surface action (in units with $\kappa=1$)
$$S=-\int dt_E L_{surf}=- A_{surf} \vert_{\beta=2 \pi}  \quad,  \eqno(16)$$ 
where integration over the Euclidean time gives a factor of $\beta=1/T=2 \pi$. This means that $E_R$ can also be expressed as
$E_R=-L_{surf}$ which agrees with eq. (15).
It is easy to see that the entropy density per horizon area in Planck units is 
$${S \over {(A/G)}}={1 \over {8{\sqrt{g}}}} \partial_i(g g^{ij})={1 \over 4} \quad, \eqno(17)$$
for all Rindler--like horizons as required.

The expression for $E_R$ given in eq. (15) is not covariant. This is not surprising since black hole entropy is observer dependent. For example, an inertial observer freely falling into the black hole sees a flat metric for which eq. (15) gives zero. This is consistent with the fact that this observer does not see the horizon. On the other hand, a Rindler observer  
sees the Rindler metric with a horizon for which eq. (15) gives the correct black hole entropy.

The derivation of $E_R$ in section 2 depends only on the time evolution of the black hole, i.e. on $g_{00}$. In order to reconcile 
this fact with the holographic formula in eq. (15) we would like to express entropy in terms of the acceleration, $a$, of an observer. Entropy should be proportional to $a$ since inertial observers such as freely falling ones do not see the horizon.
If we want to express entropy as a surface integral, then we also need the integrand to be a total derivative.
$A_{surf}$ can be written in a form consistent with these demands as[\REV]
$$A_{surf}=-{1 \over {8 \pi G}} \int d^4 x \sqrt{g} \nabla_i a^i \quad, \eqno(18)$$
where the covariant derivative is $\nabla_i a^i=(g)^{-1/2} \partial_i(g^{1/2}a^i)$ and
the four--acceleration is given by $a^i=(0,{\bf a})$ with ${\bf a}={\bf \nabla} ln(\sqrt{g_{00}})$. For the metric in eq. (8), 
$a^{\rho}=1/\rho$ is the local acceleration along the radial direction that is perpendicular to the horizon surface. 
Substituting into eq. (18) we find an alternative holographic formula for $E_R$
$$E_R={1 \over {8 \pi G}} \int_{\cal H} d^2x_{\perp} \sqrt{\sigma}(g_{00})^{1/2}{\bf \nabla} ln(\sqrt{g_{00}}) \quad, \eqno(19)$$
where $\sigma_{ij}$ is the transverse metric on the horizon. As expected, $S=2 \pi E_R$.
This establishes the relation between time evolution near the horizon fixed by $g_{00}$ and the entropy, resolving the third issue raised in the previous section.
In terms of the acceleration, the entropy per horizon area in Planck units becomes
$${S \over {(A/G)}}={1 \over 4} (g_{00})^{1/2} a^i ={1 \over 4} \quad, \eqno(20)$$ 
which is completely determined by $g_{00}$.
We see that the entropy an observer sees is determined by her world--line as required. 

$E_R$ is defined as the dimensionless Rindler energy near the horizon of the black hole. However, it is well--known that in General Relativity there are different notions of energy. Consider the energy in a volume $V$ defined by
$$E=\int_V d^3x \sqrt{g}(T_{\mu \nu}  -{1 \over 2}T g_{\mu \nu})u^{\mu} u^{\nu} \quad. \eqno(21)$$
$E$ is (one half of) the source of gravitational acceleration since for an ideal fluid we get 
$2(T_{\mu \nu}-(1/2)g_{\mu \nu}) u^{\mu} u^{\nu}=\rho+3p$. 
$E_R$ is exactly $E$ in
units with $\kappa=1$. This can be shown by using the differential geometric identity[\REV]
$$R_{\mu \nu} u^{\mu} u^{\nu}=\nabla_i(Ku^i+a^i)-K_{\mu \nu}K^{\mu \nu}+K^{\mu}_{\mu}K^{\nu}_{\nu} \quad. \eqno(22)$$
In space--times with $K_{\mu \nu}=0$, substituting eq. (22) into the definition of entropy in eq. (18), using Einstein's equation 
and the fact that time integration is just multiplication by $2 \pi$ (when $\kappa=1$) we find that $A_{surf}(\beta=2 \pi)=2 \pi E$.
Thus $E=E_R$.

Finally, we would like to comment on the relation between the surface action in eq. (12) and the Gibbons--Hawking term, $A_{GH}$,
that is added to the Einstein--Hilbert action in Euclidean gravity[\GIB] in order to make
it invariant under diffeomorphisms that vanish on the boundary. $A_{GH}$ is given by
$$A_{GH}={1 \over {8 \pi G}} \int d^4 x \sqrt{g} \nabla_c(K n^c) \quad, \eqno(23)$$
where $K$ is the trace of the extrinsic curvature and $n^c$ denotes the direction normal to the horizon. In general, $A_{surf}$ does
not coincide with $A_{GH}$[\PGH]. However, if the boundary (horizon) is given by $x^i=const.$ and the metric has no off--diagonal terms with respect to $x^i$, then $A_{surf}= A_{GH}$. This is exactly the case for the class of metrics that describe black holes with Rindler--like near horizon geometries, i.e. those given by eq. (8). 


\bigskip
\centerline{\bf 4. Rindler Energy and the AdS/CFT Correspondence}
\medskip

We saw above that $E_R$ is holographic in the sense that it is given by a surface integral over the horizon. Another notion of holography is realized by the AdS/CFT correspondence in which gravity in the AdS bulk is dual to a CFT on the
boundary. It would be interesting to find
the dimensionless energy in the boundary CFT that corresponds to $E_R$ in the bulk that is obtained from the near horizon of an AdS black hole.

Consider a black hole in $D=d+2$ dimensional AdS space given by the metric[\BIR]
$$ds^2_{d+2}=-f(r) dt^2 +f(r)^{-1} dr^2+r^2 d\Omega^2_d \quad, \eqno(24)$$
where 
$$f(r)=1+{r^2 \over L^2}-{{\omega_d M} \over {r^{d-1}}} \qquad \omega_d={{16 \pi G} \over {d V_d}} \quad. \eqno(25)$$
Here $L$ is the radius of curvature of the AdS space given by $L^2=-d(d+1)/2 \Lambda$, $G$ is the $D$-dimensional Newton constant,
$M$ is the mass of the black hole and $V_d$ is the unit volume in $d$ dimensions.
The black hole radius $R$ is the solution to
$$1+{R^2 \over L^2}-{{\omega_d M} \over {R^{d-1}}}=0 \quad. \eqno(26)$$
The Hawking temperature of the black hole is given by
$$T_H={R \over {4 \pi L^2}}\left[(d+1)+(d-1){L^2 \over R^2}\right] \quad. \eqno(27)$$
We now take the near horizon limit with $r=R+y$ and $y<<R$. 
Following the procedure in section 2 find the near horizon metric
$$ds^2=-{F(R)^2 \over 4} \rho^2 dt^2+ d\rho^2+ R^2d\Omega^2_d \quad, \eqno(28)$$
where 
$$F(R)=(d+1){R \over L^2}+(d-1){1 \over R} \quad, \eqno(29)$$
and the proper distance to the horizon is 
$$\rho=F(R)^{-1/2} \int {dy \over \sqrt{y}}=2 \sqrt{{y \over F(R)}} \quad. \eqno(30)$$
The dimensionless Rindler time is defined as $\tau_R=(F(R)/2) t$. The dimensionless Rindler energy obtained by the procedure
in section 2 gives the correct black hole entropy 
$$S=2 \pi E_R={{V_d R^d} \over {4 G}} \quad. \eqno(31)$$
In order to compute the entropy, we could, alternatively, have used eq. (15) or (19) with the metric in eq. (8) which, of course, 
gives the same result.

Using the AdS/CFT correspondence we would like to find the dimensionless energy in the boundary CFT that is analogous to $E_R$. Since the boundary CFT is conformal we can rescale its dimensions (and therefore its energy) arbitrarily. Following ref. [\VER] we rescale the boundary dimensions so that the radius of the boundary sphere is $R$. As a result, the boundary time is related to the AdS time by
$t_{CFT}=(R/L)t$. The AdS black hole is described by a state of the boundary CFT with total energy[\VER]
$$E_{CFT}={1 \over \omega_d}\left({R^d \over L}+LR^{d-2}\right) \quad. \eqno(32)$$
The energy is a sum of two terms; $E_{CFT}=E_E+E_C$ where $E_E$ and $E_C$ are the extensive energy of a gas and the Casimir energy respectively which are given by
$$E_E={1 \over \omega_d}{R^d \over L} \qquad E_C={1 \over \omega_d} LR^{d-2} \quad. \eqno(33)$$

Now, we can use the method described in section 2 for the CFT with the above energy and temperature given by $T_{CFT}=(L/R)T_H$.
We can rescale (a dimensionless) time so that $T_{CFT}=1/2 \pi$ and use eq. (5) to find the dimensionless energy conjugate to the
dimensionless time. {\footnote1{This should not be called Rindler time since the boundary is on a sphere $S^d$ and not Rindler space.}}
Since both $E_{CFT}$ and $T_{CFT}$ are rescaled by a factor of $L/R$ relative to their bulk counterparts
the procedure works the exactly same way as it does for the bulk black hole giving the same answer for entropy. However, this is not very illuminating since it does not teach us anything new about the boundary CFT.

On the other hand, in the CFT, entropy is given by the Cardy-Verlinde formula[\VER]
$$S_{CFT}={{4 \pi} \over d} R \sqrt{E_C(E_{CFT}-E_C)} \quad. \eqno(34)$$
Using the definitions of $E_{CFT}$ and $E_C$ it is easy to see that $S_{CFT}=(4 \pi L/d) E_E$. As a result, we find[\MED] 
$$E_R={{2L} \over d} E_E \quad. \eqno(35)$$
The holographic counterpart of $E_R$ is (up to the factor (2/d)) the product of the AdS radius $L$ and the extrinsic component of the CFT energy, $E_E$. The length $L$ is not part of the description of the boundary CFT since the boundary sphere has radius $R$. We can express it in terms of the CFT variables as $L=R \sqrt{(E_C/E_E)}$. 

For very large black holes with $R>>L$ we get $E_R=(2L/d) E_{CFT}$ since now the energy is dominated by the extensive part. Moreover, in this limit a rescaling by $L$ brings the CFT temperature to $T_{CFT}=(d+1)/4 \pi$. The rescaling on the boundary CFT becomes very similar to that in the bulk in which $T_H$ is rescaled to be $1/2 \pi$.

The Cardy-Verlinde formula in eq. (34) is very suggestive. Comparing it to the usual Cardy formula for two dimensional CFTs we find that the central charge and level of the CFT state that corresponds to the black hole can be expressed as[\VER]
$$ c=24 R E_C \qquad  L_0=R E_{CFT} \quad. \eqno(36)$$
From eqs. (33) and (36) we find that for the boundary CFT states that are holographic to AdS black holes, the ratio $L_0/c$ is fixed
by the ratio of the bulk variables $R/L$ as 
$${L_0 \over c}={1 \over 24} \left(1+{R^2 \over L^2} \right) \quad. \eqno(37)$$
In terms of the central charge $c$ and the level $L_0$ of the CFT state $E_R$ is given by
$$E_R={{2L} \over d}E_E={{2L} \over d}{(L_0-c/24) \over R} \quad, \eqno(38)$$
where $L$ can be expressed in terms of the CFT variables as $L=R\sqrt{24 L_0/c-1}$.

Above, we found that $E_R$ is given in terms of $E_E$, the extrinsic component of $E_{CFT}$. From eq. (34), this follows simply because of the relation $E_C=(L^2/R^2)E_E$. However the choice of $E_E$ is somewhat arbitrary. We can also express $E_R$ in terms of the boundary Casimir energy, $E_C$, as $E_R=(2R^2/dL) E_C$ instead of eq. (35). This provides an alternative expression for the holographic counterpart for $E_R$ in the boundary CFT.

\bigskip
\centerline{\bf 5. Holography and $E_R$ in Generalized Theories of Gravity}
\medskip

We saw that, in General Relativity, the Einstein--Hilbert action can be written as a sum of a bulk and a surface term.
The surface Hamiltonian obtained from the surface action, when evaluated on the horizon, is precisely $E_R$.
Then, $E_R$ is given by a holographic expression as an integral over the horizon surface. Black hole entropy is 
$S=2\pi H \vert_{\cal H}=2\pi E_R$. Alternatively, $S=-A_{surf}$ and $E_R$ is given by (the negative of) the surface Lagrangian.

Since black hole entropy in generalized theories of gravity is also given by $2 \pi E_R$, it is important to find out if the same ideas can be applied beyond General Relativity. For this purpose consider a four--dimensional theory of gravity with the Lagrangian density
${\cal L}= Q_a^{bcd} R^a_{bcd}$. This includes actions with any power of the Riemann tensor with up to two derivatives of the metric.
In these theories, as in General Relativity, the Lagrangian density can be divided into a bulk and surface term[\REV,\DIV]
$${\cal L}_{bulk}=2Q_a^{bcd} \Gamma^a_{dk} \Gamma^k_{bc} \qquad {\cal L}_{surf}=2 \partial_c(\sqrt{g}Q_a^{bcd} \Gamma^a_{bd}) \quad. \eqno(39)$$
As in General Relativity, ${\cal L}_{bulk}$ is quadratic in the first derivatives whereas ${\cal L}_{surf}$ contains second derivatives. The surface action evaluated on the horizon can be written as
$$A_{sur}=2 \int d^4x \partial_c(\sqrt{g}Q^{abcd} \partial_b g_{ad}) \quad. \eqno(40)$$
which is the generalization of eq. (11).
Integrating over the Euclidean time with period $\beta=2 \pi$ and using Gauss' law we get
$$A_{sur}=4 \pi \int_{\cal H} d^2x_{\perp} \sqrt{g} Q^{abr0} \partial_b g_{a0} \quad, \eqno(41)$$
where $\cal H$ is the horizon surface. The dimensionless Rindler energy is given by $E_R=H_{surf}\vert_{\cal H}=A_{surf}/2 \pi$.
For $A_{surf}$ to give the correct black hole entropy and therefore $E_R$ in any theory of gravity, eq. (41)
has to be identical to Wald entropy[\LAST] given by
$$S_{Wald}=2 \pi Q=2 \pi \int_{\cal H}d^2x_{\perp} Y^{abcd}\epsilon_{ab} \epsilon_{cd}  \quad, \eqno(42)$$
where $Q$ is Wald's Noether charge, $\epsilon_{ij}$ are the binormals to the horizon and $Y^{abcd}=(\delta L / \delta R_{abcd})$.

$A_{surf}$ is identical to Wald entropy, or $E_R=Q$, only when $Y^{abcd}=Q^{abcd}$. In general, this is not the case and therefore
the surface action in eq. (41) does not equal black hole entropy. For example, consider the Lanczos--Lovelock gravity with
the Lagrangian density[\LOV,\LAN]
$${\cal L}=\sum_m{\cal L}_m=\sum_{k=1}^{2m}{1 \over {16 \pi G_D}} \delta^{1357\ldots 2k-1}_{2468\ldots 2k} R^{24}_{13} R^{68}_{57} \ldots R^{2k-2~ 2k}_{2k-3~ 2k-1}
\quad, \eqno(43)$$
where $m$ is the order. Since ${\cal L}_m$ is a homogeneous function of $R^{ab}_{bc}$ of degree $m$, at any order $m$, we get $Y^{abcd}=m Q^{abcd}$. As a result, the surface action obtained from eq. (41) satisfies $mA_{surf}=S_{Wald}$[\MSUR]. Thus, the surface action in eq. (41) gives the entropy only for $m=1$ which corresponds to General Relativity. For $m >1$, $A_{surf}$ is a fraction ($1/m$ times) of the entropy. For Lanczos--Lovelock theories with more than one term in eq. (43), $A_{surf}$ is not proportional to entropy due to the different multiplicative factors.
For example, in Einstein--Gauss--Bonnet gravity which corresponds to the sum of the $m=1$ and $m=2$ terms, $A_{surf}$ is not proportional to
black hole entropy which is given by $2 \pi E_R$[\GB].

The situation is even worse for some theories of gravity, such as $D=2+1$ topologically massive gravity, in which black hole entropy depends on parameters that do not appear in the metric[\GRU]. Since the holographic formula in eq. (15) depends only the metric, in these theories it cannot possibly give the entropy.
{\footnote2{I would like to thank D. Grumiller for raising this issue and an illuminating correspondence.}} On the other hand, in these theories as in all theories of gravity, entropy is still equal to $2 \pi E_R$. This can be explained by noting that
the original definition of $E_R$ in terms of the Poisson bracket in eq. (5) involves the mass $M$ that the depends on these parameters. Therefore, the definition of $E_R$ in terms of the Poisson bracket in eq. (5) is more general than the holographic
formula in eq. (15).

We found that even though for a large class of generalized theories of gravity, the action can be written as a sum of a bulk and a surface term, the surface action does not correspond to black hole entropy. Therefore, $E_R$ is not given by the surface Hamiltonian evaluated on the horizon. Of course, since $E_R$ is exactly Wald's Noether charge $Q=S_{Wald}/2 \pi$, we can always use eq. (42)
to write $E_R$ as a surface integral. However, in general, $Q$ is not the surface Hamiltonian. We conclude that even though
$E_R$ can always be written as a surface integral as in eq. (42) it is not always given by $H_{surf}$. It would be interesting to find the relation between the Wald's Noether charge and $H_{surf}$ is generalized theories of gravity.

\bigskip
\centerline{\bf 6. Conclusions and Discussion}
\medskip

In this paper, we obtained a formula for the dimensionless Rindler energy, $E_R$, which gives black hole entropy as
$S=2 \pi E_R$ in all theories of gravity. Eq. (15) shows that $E_R$ is given by a surface integral over the horizon making its holographic nature manifest. $E_R$ can also be expressed as a surface integral of the local acceleration as in eq. (19). This formula only depends
on $g_{00}$ and shows the relation between black hole entropy and time evolution in the near horizon region. These expressions
for $E_R$ are not covariant; i.e. they are observer dependent as required for black hole entropy.
Using the AdS/CFT correspondence, we also obtained the holographic counterpart of $E_R$ in the boundary CFT. This is given by the product of the AdS radius and the extensive part of energy in the CFT. 

The holographic expressions for $E_R$ are obtained by using a special property of the Einstein--Hilbert action, i.e. that it can be divided into a bulk and a surface term. Then, it is easy to show that the surface Hamiltonian evaluated on the horizon is exactly $E_R$. Alternatively, $E_R$ is given by (the negative of) the surface Lagrangian. The generalization of these results beyond
General Relativity is not straightforward.
We found that even though for a wide class of theories of gravity the action can be divided into a bulk and a surface part, the surface Hamiltonian does not equal $E_R$. On the other hand, since $E_R=Q$ in all theories of gravity we can use eq. (42) to write it
as a surface integral. Therefore, $E_R$ is holographic in all theories of gravity but its relation to the surface Hamiltonian
is not clear in general.

In this paper, we did not address the most important question about $E_R$, namely the nature of the fundamental degrees of freedom it counts. The dimensionless Rindler energy is the generator of time translations near the horizon with the normalization $T_H=1/2 \pi$. 
Therefore, it is related to the evolution of the black hole in time. On the other hand, $E_R$ is given by a surface integral over the horizon and is therefore related to horizon degrees of freedom, The expression for $E_R$ in terms of the local acceleration, e.g.
eq. (19), involves both $g_{00}$ and a surface integral over the horizon and therefore may be a first hint about the relation between these two different descriptions of Rindler energy.
In ref. [\LAST], we speculated on the nature of the degrees of freedom that $E_R$ counts but needless to say this remains an important open question.

\bigskip
\centerline{\bf Acknowledgments}

I would like to thank the Stanford Institute for Theoretical Physics for hospitality.

\vfill

\refout

\end
\bye